\documentclass[aps,prl,twocolumn,showpacs,floatfix,superscriptaddress,letterpaper,longbibliography]{revtex4-1}
\usepackage{graphicx}
\usepackage{bm}
\usepackage{amsmath,amsfonts}
\usepackage{amsbsy}
\usepackage[utf8]{inputenc}
\DeclareUnicodeCharacter{03BC}{\mbox{$\mu$}}
\DeclareUnicodeCharacter{2009}{ }

\begin{document}

\title{Hybrid sound modes in one-dimensional quantum liquids}

\author{K. A. Matveev}

\affiliation{Materials Science Division, Argonne National Laboratory,
  Argonne, Illinois 60439, USA}

\author{A. V. Andreev}

\affiliation{Department of Physics, University of Washington, Seattle,
  Washington 98195, USA}

\date{March 28, 2018}

\begin{abstract}
  We study sound in a single-channel one-dimensional quantum liquid.
  In contrast to classical fluids, instead of a single sound mode we
  find two modes of density oscillations.  The speeds at which these
  two sound modes propagate are nearly equal, with the difference that
  scales linearly with the small temperature of the system.  The two
  sound modes emerge as hybrids of the first and second sounds, and
  combine oscillations of both density and entropy of the liquid.
\end{abstract}
\maketitle

Small oscillations of density propagate through fluids in the form of
sound waves.  At long wavelengths the compression and rarefaction of
the fluid are adiabatic, and the speed of sound is determined by the
adiabatic compressibility,
\begin{equation}
  \label{eq:speed_of_sound}
  v_1=\sqrt{\left(\frac{\partial P}{\partial \rho}\right)_\sigma},
\end{equation}
where $P$, $\rho$, and $\sigma$ are the pressure, mass density, and
entropy per unit mass of the fluid, respectively.  Importantly, there
is only one type of sound in the system. 

A well-known exception to the above picture is the superfluid $^4$He.
It can be viewed as consisting of two fluids: the superfluid component
and the normal one \cite{landau_theory_1941,
  khalatnikov_introduction_2000}.  The latter is the gas of elementary
excitations, which can move with respect to the center of mass of the
fluid even in thermodynamic equilibrium.  As a result, the superfluid
$^4$He supports two types of sound waves.  The first sound is
predominantly an oscillation of the fluid density $\rho$.  It is
similar to the sound in ordinary fluids; in the low temperature limit
its velocity is given by Eq.~(\ref{eq:speed_of_sound}).  The second
sound is a wave of entropy, which is for the most part decoupled from
the oscillations of density.  Its speed is
\begin{equation}
  \label{eq:v_2}
  v_2=\sqrt{\frac{\rho_s\sigma^2}{\rho_n(\partial\sigma/\partial T)_\rho}},
\end{equation}
where $\rho_n$ and $\rho_s=\rho-\rho_n$ are the densities of the
normal and superfluid components, respectively.  The velocities of the
two sound waves are different.  In the most important case of low
temperature $T$, when the excitations are phonons with linear
dispersion, $v_1/v_2=\sqrt{3}$.

The calculation of $v_1/v_2$ can be easily generalized to the case of
space of any dimension $d$, resulting in $v_1/v_2=\sqrt{d}$ at
$T\to0$.  This raises an interesting question regarding the fate of
the first and second sounds in one dimension, where $v_1=v_2$.
Indeed, Eqs.~(\ref{eq:speed_of_sound}) and (\ref{eq:v_2}) give the
speeds of the two sounds only if $v_1\neq v_2$.  Their derivation
\cite{landau_theory_1941, khalatnikov_introduction_2000} reveals the
existence of coupling between the first and second sounds, resulting
in corrections to Eqs.~(\ref{eq:speed_of_sound}) and (\ref{eq:v_2}),
which disappear when $T/(v_1-v_2)\to0$.  In one spatial dimension,
where $v_1-v_2\to0$ at $T\to0$, the coupling between the first and
second sounds and the physical nature of the resulting acoustic modes
are open questions.

One-dimensional quantum fluids do not undergo superfluid transition.
Nevertheless, many of their properties mirror those of superfluid
$^4$He.  In particular, their low energy excitations, typically
described in the framework of the Tomonaga-Luttinger liquid theory
\cite{tomonaga_remarks_1950, haldane_luttinger_1981}, are bosons,
which are analogous to phonons in $^4$He.  Because of the absence of
superfluidity, in thermodynamic equilibrium the velocity of the gas of
excitations of a one-dimensional quantum fluid must equal that of the
center of mass of the system.  However, the relaxation processes
leading to equilibration of these two velocities are exponentially
slow, with the corresponding relaxation rate scaling as
$\tau^{-1}\propto e^{-D/T}$ \cite{matveev_equilibration_2010,
  matveev_equilibration_2012}.  (Here $D$ is the energy bandwidth of
the model.)  This is in contrast to the much faster rate $\tau_{\rm
  ex}^{-1}$ of collisions of the excitations with each other, which
scales as a power of temperature \cite{imambekov_one-dimensional_2012,
  arzamasovs_kinetics_2014, protopopov_relaxation_2014,
  ristivojevic_relaxation_2013}.  Thus, in a broad range of
frequencies
\begin{equation}
  \label{eq:frequency_range}
  \tau^{-1} \ll \omega \ll \tau_{\rm ex}^{-1},
\end{equation}
the excitations form a gas that can move as a whole with respect to
the fluid.  In this regime the one-dimensional system behaves as a
superfluid and supports both first and second sounds
\cite{matveev_second_2017}.

It is important to point out that the theory of one-dimensional
quantum fluids developed in Ref.~\cite{matveev_second_2017} focused on
systems of fermions with spins, which support excitations of two
types, corresponding to the charge and spin degrees of freedom.
Because they propagate at different speeds, the aforementioned result
$v_1/v_2=\sqrt{d}$ does not apply, and in this system $v_1\neq v_2$.
On the other hand, a system of spinless bosons (or that of fermions
with spins polarized by an external magnetic field) has only a single
channel of bosonic excitations propagating with velocity $v$, which
depends on the strength of interactions between the constituent
particles.  For such a liquid, Eqs.~(\ref{eq:speed_of_sound}) and
(\ref{eq:v_2}) predict $v_1=v_2=v$ at $T\to0$.  This leads to strong
enhancement of coupling between the first and second sounds, which
changes dramatically the nature of the acoustic modes in the system.
The study of this phenomenon is the main goal of this paper.

We consider a Galilean invariant system of identical interacting
spinless particles of mass $m$.  The low energy properties of such
quantum systems are described by the Luttinger liquid theory
\cite{haldane_luttinger_1981}.  In this approach the excitations of
the system are bosons with momenta $k$ that are multiples of
$2\pi\hbar/L$, where $L$ is the system size, and periodic boundary
conditions are assumed.  The energy $\varepsilon$ and momentum $p$ of
the system per unit length may be expressed in terms of the occupation
numbers $N_k$ of the bosonic states and two additional zero-mode
variables: the total number of particles $N$ and the integer quantum
number $J$ \cite{onefootnote},
\begin{eqnarray}
  \label{eq:E}
  \varepsilon&=&\frac{mv^2}{2N_0L}(N-N_0)^2 + \frac{\pi^2\hbar^2}{2mL^3} NJ^2
    +\frac{1}{L}\sum_k\epsilon_k N_k,
\\
  \label{eq:p}
  p&=&\frac{\pi\hbar}{L^2}NJ+\frac{1}{L}\sum_kk N_k.
\end{eqnarray}
In Eq.~(\ref{eq:E}) $v$ is the speed of sound at zero temperature,
$\epsilon_k$ is the energy of the bosonic excitation with momentum
$k$, and the number of particles is assumed to be near some reference
value $N_0$.

The first term in the expression (\ref{eq:p}) for the momentum density
accounts for the fact that even in the absence of bosonic excitations
the system can move as a whole.  Instead of the quantum number $J$ it
will be convenient to quantify the momentum associated with this
motion by the velocity
\begin{equation}
  \label{eq:u_0}
  u_0=\frac{\pi\hbar J}{mL}.
\end{equation}
The bosonic excitations in the Luttinger liquid theory are usually
assumed to propagate with the velocity $v$, resulting in the energy
spectrum $\epsilon_k=v|k|$.  It is important to note that this is
appropriate only at $J=0$, i.e., in the reference frame moving with
velocity $u_0$.  The energy spectrum in the stationary frame,
\begin{equation}
  \label{eq:epsilon}
  \epsilon_k=v|k|+u_0k,
\end{equation}
can be obtained by the Galilean transformation.

To study sound modes, one has to account for the relaxation processes
in the system.  The collisions between the bosonic excitations occur
at the relatively short time scale $\tau_{\rm ex}$.  Thus at
frequencies $\omega\ll\tau_{\rm ex}^{-1}$ the distribution function
has the equilibrium form
\begin{equation}
  \label{eq:N_k}
  N_k=\frac{1}{e^{(\epsilon_k-u_{\rm ex}k)/T}-1}.
\end{equation}
Collisions of the bosonic excitations conserve their total momentum,
given by the second term in the right-hand side of Eq.~(\ref{eq:p}).
As a result, the equilibrium state can be realized with an arbitrary
value of the velocity $u_{\rm ex}$.  At the much longer time scale
$\tau$ the bosonic excitations exchange momentum with the zero mode
$J$, and only the total momentum (\ref{eq:p}) is conserved.  This
gives rise to the slow relaxation $u_{\rm ex}-u_0\to0$.

In the frequency range (\ref{eq:frequency_range}), in addition to the
number of particles, momentum, and energy, the quantum number $J$ is
also a conserved quantity.  As a result, the state of the fluid is
characterized by two velocities, $u_0$ and $u_{\rm ex}$.  This enables
one to develop a two-fluid hydrodynamic description of the system
\cite{matveev_second_2017} analogous to that of superfluid $^4$He
\cite{landau_theory_1941, khalatnikov_introduction_2000}.  In this
theory, $u_0$ and $u_{\rm ex}$ play the roles of the velocities of the
superfluid and normal components, respectively.

Similar to the superfluid $^4$He, one-dimensional quantum fluids
support two sound modes.  Equation (\ref{eq:speed_of_sound}) is the
standard expression for the speed of sound in a fluid.  Because the
parameter $v$ in the Luttinger liquid theory is the speed of sound at
zero temperature, by definition $v_1\to v$ at $T\to0$.  To obtain the
speed of the second sound, we need to determine the values of the
normal and superfluid densities $\rho_n$ and $\rho_s$.  To this end we
evaluate the momentum density of the system at low temperatures using
Eqs.~(\ref{eq:p}) and (\ref{eq:N_k}) and find it to have the expected
form $p=\rho_s u_0+\rho_n u_{\rm ex}$, where
\begin{equation}
  \label{eq:rhos}
  \rho_s=\rho-\rho_n,
\quad
  \rho_n=\rho\chi,
\quad
  \chi=\frac{\pi T^2}{3\hbar \rho v^3}.
\end{equation}
We then evaluate the entropy per
unit mass $\sigma$ in the equilibrium state with $u_0=u_{\rm ex}=0$
and find
\begin{equation}
  \label{eq:sigma}
  \sigma=\frac{\pi T}{3\hbar\rho v}.
\end{equation}
Substituting these expressions into Eq.~(\ref{eq:v_2}), in the limit
$T\to0$ we obtain the expected result $v_2=v$.  

The equality of the speeds of the first and second sounds enhances the
effect of interaction between them.  To study the consequences of this
interaction, we use the equations of the two-fluid hydrodynamics,
which express the conservation laws of the number of particles,
momentum, energy, and $J$.  These four equations can be either derived
microscopically, following the procedure described in
Ref.~\cite{matveev_second_2017}, or simply obtained by adapting the
hydrodynamic equations for superfluid $^4$He \cite{landau_theory_1941,
  khalatnikov_introduction_2000} to one spatial dimension.  The state
of the fluid is described by four dynamic variables, e.g.~$\rho$,
$\sigma$, $u_0$, and $u_{\rm ex}$.  In the frequency range
(\ref{eq:frequency_range}) one can neglect dissipative processes.  In
this case it is possible to reduce the system of four first-order
differential equations of two-fluid hydrodynamics to two second-order
equations by excluding the two velocities.  This yields
\cite{landau_theory_1941, khalatnikov_introduction_2000}
\begin{subequations}
  \label{eq:sound_eqns}
\begin{eqnarray}
  \label{eq:rho_eqn}
  \partial_t^2\rho &=& \partial_x^2 P,
\\[1ex]
  \label{eq:sigma_eqn}
  \partial_t^2\sigma &=& \frac{\rho_s}{\rho_n}\sigma^2\partial_x^2 T.
\end{eqnarray}
\end{subequations}
In the theory of superfluid $^4$He, the sound modes have been studied
\cite{landau_theory_1941, khalatnikov_introduction_2000} by using $T$
and $P$ as independent variables in Eq.~(\ref{eq:sound_eqns}).  In
order to elucidate the nature of the sound modes in one-dimensional
quantum systems, it will be more convenient to choose $\rho$ and
$\sigma$ as independent variables, and treat pressure and temperature
as their functions, $P(\rho,\sigma)$ and $T(\rho,\sigma)$.

We now linearize Eq.~(\ref{eq:sound_eqns}) in small deviations of
$\rho$ and $\sigma$ from their equilibrium values.  Substituting
the deviations in the form $\delta\rho\cos [q(x-ut)]$ and
$\delta\sigma\cos [q(x-ut)]$ into Eq.~(\ref{eq:sound_eqns}), we obtain
\begin{subequations}
  \label{eq:sound_eqns_linearized}
\begin{eqnarray}
  \label{eq:rho_eqn_linearized}
  (u^2-v_1^2)\,\delta\rho-A\,\delta\sigma&=&0,
\\[1ex]
  \label{eq:sigma_eqn_linearized}
  -B\,\delta\rho+(u^2-v_2^2)\,\delta\sigma&=&0.
\end{eqnarray}
\end{subequations}
Here $v_1$ and $v_2$ are given by Eqs.~(\ref{eq:speed_of_sound}) and
(\ref{eq:v_2}), respectively, and 
\begin{equation}
  \label{eq:B}
  A=\bigg(\frac{\partial P}{\partial\sigma}\bigg)_{\!\rho},
\qquad
  B=\frac{\rho_s}{\rho_n}\,\sigma^2
           \bigg(\frac{\partial T}{\partial\rho}\bigg)_{\!\sigma}.
\end{equation}
The system of linear equations (\ref{eq:sound_eqns_linearized}) has
nontrivial solutions at
\begin{equation}
  \label{eq:u^2}
  u^2=\frac{v_1^2+v_2^2}{2}
    \pm\frac12\sqrt{(v_1^2-v_2^2)^{2}+4AB}.
\end{equation}
This expression gives the speeds of the two sound modes in a
one-dimensional quantum liquid.

Equation (\ref{eq:u^2}) also applies to superfluid $^4$He.  In that
case, it was shown \cite{landau_theory_1941,
  khalatnikov_introduction_2000} that $AB\to0$ at $T\to0$, while
$v_1-v_2$ approaches a nonzero value.  Therefore at low temperatures
$u$ becomes either $v_1$ or $v_2$, corresponding to the first or
second sound, respectively.  This is also the case for a
one-dimensional system of fermions with spin considered in
Ref.~\cite{matveev_second_2017}.  In contrast, for a spinless
one-dimensional quantum liquid, $v_1-v_2\to0$ at $T\to0$, and the
analysis of sound modes requires accounting for finite temperature
corrections. 

We start by determining the temperature dependence of the quantities
$A$, $B$, $v_1^2$, and $v_2^2$ in Eq.~(\ref{eq:u^2}).  Using the
thermodynamic Maxwell's relation $(\partial P/\partial\sigma)_\rho =
\rho^2(\partial T/\partial\rho)_\sigma$ and Eqs.~(\ref{eq:rhos}),
(\ref{eq:sigma}), and (\ref{eq:B}), we find
\begin{equation}
  \label{eq:AB}
  A=\frac{\rho T}{v}\partial_\rho(\rho v),
\quad
  B=\frac{v^3\chi}{\rho T}\partial_\rho(\rho v).
\end{equation}
Both $A$ and $B$ scale linearly with the temperature.

To leading order in $T$, the temperature dependence of the free energy
of a Luttinger liquid can be easily obtained from Eq.~(\ref{eq:E}).
This is sufficient to evaluate the correction to the pressure $P$,
which is proportional to $T^2$.  As a result, the correction to speed
of sound (\ref{eq:speed_of_sound}) appears in the same order,
\begin{equation}
  \label{eq:v_1_correction}
  v_1=v+\frac{\pi T^2}{12\hbar\rho v^3}\,\partial_\rho^2(\rho^2 v).
\end{equation}
This approach yields only the leading temperature dependence for the
entropy (\ref{eq:sigma}) and is thus insufficient to obtain the
temperature-dependent correction to the velocity $v_2$ from
Eq.~(\ref{eq:v_2}).

On the other hand, corrections to $v_2$ can be obtained in some
important cases.  For non-interacting spinless fermions, the
Sommerfeld expansion of the free energy involves only even powers of
$T$, resulting in
\begin{equation}
  \label{eq:v_2_fermions}
  v_2=v-\frac{73\pi T^2}{30\hbar\rho v^2}.
\end{equation}
For the model of one-dimensional bosons with short-range repulsion
\cite{lieb_exact_1963}, the thermodynamics can be studied by means of
the thermodynamic Bethe ansatz \cite{yang_thermodynamics_1969}.  The
Sommerfeld expansion of the free energy of this system also contains
only even powers of $T$ \cite{guan_ferromagnetic_2007,
  guan_polylogs_2011}, yielding a correction to $v_2$ that scales as
$T^2$.  These examples strongly suggest that the quadratic temperature
dependence of the corrections to $v_2$ is a generic property of
one-dimensional quantum liquids.  For systems with interactions that
decay with distance faster that $1/x^3$, this property can be
demonstrated by a careful treatment of irrelevant perturbations to the
Hamiltonian of the Luttinger liquid \cite{unpublished}.

Taking into account the temperature dependences $v_1^2-v_2^2\propto
T^2$ and $AB\propto T^2$, we approximate the square root in
Eq.~(\ref{eq:u^2}) by $\sqrt{4AB}$.  This results in linear in $T$
splitting of the sound velocities,
\begin{equation}
  \label{eq:u_pm}
  u_\pm=v\pm\frac{\sqrt{\chi}}{2}\partial_\rho(\rho v),
\end{equation}
where we applied Eq.~(\ref{eq:AB}).  

Importantly, the physical character of the two sound modes that
propagate at the speeds (\ref{eq:u_pm}) is very different from those
of the first and second sounds.  To demonstrate this, we obtain the
relative magnitudes of the oscillations of $\rho$ and $\sigma$ by
substituting $u_\pm$ into Eq.~(\ref{eq:sound_eqns_linearized}) and
find $\delta\rho/\delta\sigma=\pm\sqrt{A/B}$.  It is instructive to
express this result in terms of variations of the particle density
$n=\rho/m$ and entropy density $s=\sigma\rho$,
\begin{equation}
  \label{eq:delta_s}
  \frac{\delta s}{\delta n}=\pm\frac{\pi}{\sqrt{3K}}.
\end{equation}
Here $K=\pi\hbar n/mv$ is the usual Luttinger liquid parameter.

Equation (\ref{eq:delta_s}) shows that the ratio of the amplitudes of
oscillations of $s$ and $n$ remains finite at $T\to0$.  This is in
contrast with the usual first sound, which is predominantly a density
wave, with $\delta s/\delta n\propto T$, and the second sound---the
wave of entropy, with $\delta n/\delta s\propto T$.  Thus the two
sound modes in one-dimensional quantum liquids are qualitatively
different from the first and second sounds, but combine the essential
characteristics of both.  This hybrid nature of sound in one dimension
is our main result.

To illustrate the unusual properties of the hybrid sound modes, let us
discuss the evolution of a local initial perturbation of density
$\delta n(x)=f(x)$, assuming that the entropy is not perturbed,
$\delta s(x)=0$.  In a classical fluid, the perturbation propagates in
both directions at the speed of sound, $\delta n(x,t)
=\frac12[f(x-v_1t) +f(x+v_1t)]$, while to first approximation the
entropy density remains undisturbed.  The evolution of density
perturbations in the one-dimensional system of fermions with spin is
qualitatively similar.  The behavior of spinless quantum liquids is
dramatically different: the density perturbation splits into four
pulses,
\begin{eqnarray}
  \label{eq:n_pulses}
  \delta n(x,t) &=&\frac14[f(x-v_+t)+f(x+v_+t)
\nonumber\\
                 &&+f(x-v_-t)+f(x+v_-t)].
\end{eqnarray}
These density pulses are accompanied by the entropy disturbance of the
form
\begin{eqnarray}
  \label{eq:s_pulses}
  \delta s(x,t) &=&\frac{\pi}{4\sqrt{3K}}[f(x-v_+t)+f(x+v_+t)
\nonumber\\
                 &&-f(x-v_-t)-f(x+v_-t)].
\end{eqnarray}
Let us now comment on the evolution of an entropy perturbation created
by a local heating of the system.  In this case all three kinds of
fluids behave very differently.  In a classical fluid the entropy
spreads diffusively, in accordance with the Fourier's law of heat
transfer.  In a quantum fluid of fermions with spin the entropy
disturbance propagates as two pulses moving in opposite directions at
the speed of the second sound.  In a spinless quantum fluid, such a
disturbance again splits into four pulses of both entropy and density,
in analogy with Eqs.~(\ref{eq:n_pulses}) and (\ref{eq:s_pulses}).

In summary, we studied sound propagation in single-channel
one-dimensional quantum liquids.  In the frequency range
(\ref{eq:frequency_range}), the system may be described by two-fluid
hydrodynamics and supports two sound modes.  In contrast to superfluid
$^4$He and one-dimensional systems with spin, the two modes are
qualitatively different from the first and second sounds, which
correspond to predominantly density or entropy waves.  Each of these
hybrid modes combines comparable in magnitude oscillations of both
density and entropy, Eq.~(\ref{eq:delta_s}).  The difference of speeds
(\ref{eq:u_pm}) of the hybrid modes scales linearly with the
temperature.

\begin{acknowledgments}

  Work at Argonne National Laboratory was supported by the
  U.S. Department of Energy, Office of Science, Materials Sciences and
  Engineering Division.  Work at the University of Washington was
  supported by the U.S.  Department of Energy Office of Science, Basic
  Energy Sciences under Award No. DE-FG02-07ER46452.

\end{acknowledgments}

\end{document}